\documentclass[journal,twocolumn,10pt]{IEEEtran} 
\usepackage{epsfig,cite}
\usepackage{graphicx}
\graphicspath{ {./figures/} }
\usepackage{tabularx}
\usepackage{caption}
\usepackage{subcaption}
\usepackage[font=small,labelfont=bf]{caption}
\usepackage[T1]{fontenc}
\usepackage{mathptmx}
\usepackage{color}
\usepackage{amssymb,amsmath,mathrsfs}
\usepackage[left=1.587cm,right=1.824cm,top=1.905cm,bottom=2.54cm]{geometry}
\usepackage{url}

\newcommand{\rev}[1]{\textcolor{black}{#1}}
\begin{document}
\title{Communication System Design using Synthetic Photoisomerizable Azobenzene-Regulated K$^+$ (SPARK) channel}
\author{
	\IEEEauthorblockN{Taha Sajjad and Andrew W. Eckford}%
    \thanks{The authors are with the Department of Electrical Engineering and Computer Science, York University, 4700 Keele Street, Toronto, Ontario, Canada M3J 1P3. Emails:  tahaatiq@cse.yorku.ca, aeckford@yorku.ca}%
	\thanks{This work was supported by a grant from the Lloyd's Register Foundation International Consortium of Nanotechnologies (LRF ICoN).}%
}

\maketitle
\begin{abstract}
 Biomolecules exhibit a remarkable property of transforming signals from their environment. This paper presents a communication system design using a light-modulated protein channel: Synthetic Photoisomerizable Azobenzene-regulated K$^+$ (SPARK). Our approach involves a comprehensive design incorporating the SPARK-based receiver, encoding methods, modulation techniques, and detection processes. 
 By analyzing the resulting communication system, we determine how different parameters influence its performance.
 Furthermore, we explore the potential design in terms of bioengineering and demonstrate that the data rate scales up with the number of receptors, indicating the possibility of achieving high-speed communication.
\end{abstract}

\section{Introduction}
Multicellular organisms receive information from the environment through many signal transduction pathways. This intricate system involves various input signals, such as changes in light intensity, shifts in chemical concentration, responses to mechanical forces, temperature variations, and other environmental alterations. This complex signal reception and interpretation is crucial for coordinating cellular actions. Different cells within the organism actively communicate this information, ensuring a harmonized response to the dynamic environmental cues.

This paper investigates the design of a light-based communication system using \rev{the Synthetic Photoisomerizable Azobenzene-Regulated K$^+$ (SPARK) channel. SPARK emerges as a versatile biomolecular component that functions as both a transducer and receiver.} 
The channel offers superior controllability, responding to light stimuli in both {\em closed} and {\em open} states \cite{fortin2011engineering}, \cite{yager2006novel}, \cite{zhang2014taking}, \cite{gorostiza2008nanoengineering}. This controllability is achieved through the precise control of azobenzene molecule isomerization. \rev{By converting optical signals into chemical signals, SPARK regulates ion flow through the cell membrane in response to light, directly influencing the membrane potential and triggering downstream signalling pathways. \textcolor{black}{While traditional molecular communication systems involve significant propagation delays due to diffusion, ion channels enable nearly instantaneous ion flow once they are open. This is supported by experimental studies, such as patch-clamp measurements, which demonstrate that ion current transitions occur immediately upon channel opening \cite{moffett2022permissive}}.} This unique characteristic of SPARK as a photoswitch \cite{beharry2011azobenzene},\cite{izquierdo2014two},\cite{zhu2018azobenzene},\cite{repina2017light}. opens up many applications in various fields, such as optogenetics, where it can precisely regulate cellular activity, allowing the detailed study of neural circuits and its behaviour \cite{chambers2008light}, \cite{berlin2020molecular},\cite{sandoz2013optogenetic}, \cite{berlin2017synapses}. 

Another promising frontier for SPARK lies in molecular computing and information analysis. Here, SPARK contributes to the design of molecular-level logic and computing. 
By harnessing these properties, researchers can explore novel approaches to information processing and computation at the molecular level, paving the way for advancements in computing technologies \cite{baroncini2015eternal}.\par
\textcolor{black}{
Unlike previous works \cite{Biomolecule}, which primarily focused on passive state transitions, our approach in this paper establishes full control over both switching states, offering a higher degree of precision in signal modulation.
A key contribution of this work is to establish the SPARK as a potential receiver in the communication system by demonstrating how the absence of external light intensity and the lack of control over the reverse transition influence system performance. Our results reveal that under these conditions, the error rates exhibit significant variations, underscoring the critical role of light-induced transitions in optimizing communication fidelity.}
By integrating the principles of biomolecular phototransduction with discrete-state Markov modelling, this work advances the potential of bio-inspired communication systems, enabling the development of high-speed, light-driven molecular networks.

 The system is modelled as a discrete-time, finite-state Markov chain with transition rates governed by the input signal \cite{ching2006markov},\cite{allen2010introduction}, \cite{vijayabaskar2017introduction}, \cite{ye2016stochastic}, \cite{chodera2014markov}. We determine the information rate of a system by applying the results of the information theory established in \cite{channelrhodopsin2018},\cite{thomas2016shannon}, validated by simulations. We also compare its performance for different chemical rates. Additionally, we validate that the findings from previous studies \cite{channelrhodopsin2018}, \cite{thomas2016shannon}, can be extended to SPARK, which suggests that the information rate increases with the growing number of molecules. 





The paper is organized as follows. In Section II, we describe our basic system and its theoretical aspects. In Section III, we present a communication model to analyze the performance of our proposed system. In Section IV, we extended our model by considering multiple receptors and presented our results in terms of data rate and probability of error in Section V. In Section VI, we conclude and discuss the prospects in this domain.

\section{Biocommunication model}

\subsection{Basic Setup}
\begin{figure}[t!]
\centering
\includegraphics[width=\columnwidth]{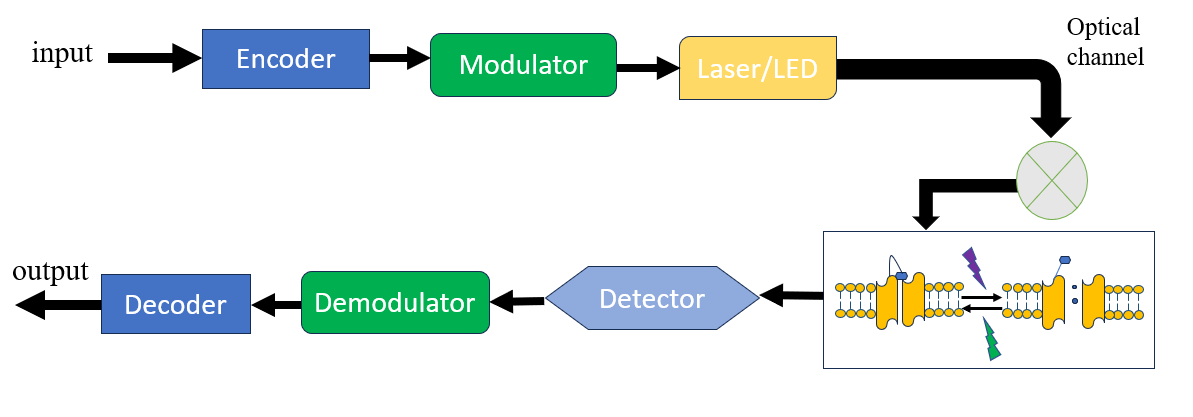}
\caption[\hspace{0.2cm}]{Block diagram of our model. The transmitter comprises of an encoder, modulator and light source, whereas the receiver side photoswitch SPARK is used followed by detector, demodulator and decoder.
Figure adapted from \cite{idris2019visible}.}
\label{fig1}
\end{figure}

The structure of our proposed model aligns closely with the one presented in \cite{khan2017visible}, including essential components such as an encoder, modulator,  light source and photo-receiver. However, we have substituted the photo-receiver with a biomolecule SPARK on the receiver side, as illustrated in Fig. \ref{fig1}. \rev{A detailed description of the components of molecular structure is provided in the following section.}

\subsection{SPARK Physical Structure and Ion Channel}
\paragraph{Ion Channel}
 \rev{Ion channels are protein structures embedded in cell membranes that allow ions (such as sodium, potassium, calcium, and chloride) to pass into or out of the cell. These channels are crucial for a variety of physiological processes, including electrical signalling in neurons, muscle contraction, regulation of cell volume, among others. Types of ion channels include voltage-gated, ligand-gated, mechanically-gated, and light-gated channels, referring to the type of stimulus that opens the channel. In this work, we will use a light-gated channel \cite{hille2012electrical},\cite{lemoine2013optical}.}\par
 \rev{The ion channel function can be modelled as a series of resting/deactivated states and active/ion flow states. In the resting state, the ion channel is closed, preventing ion flow. Upon receiving a specific stimulus, such as light or a specific frequency in the visible spectrum, the ion channel undergoes a conformational change that opens the channel, allowing ions to move through the open channel according to their electrochemical gradient.}
 \paragraph{Azobenzene}Azobenzene is a chemical compound of two benzene rings connected by a central nitrogen-nitrogen (N=N) double bond; the linkage is termed {\em azo} \cite{xue2021fluorescent} Fig. \ref{figazo}.
 The remarkable common characteristic shared by azobenzene molecules (referred to as azos) is their efficient and precise photochemical isomerization when the molecule absorbs a photon: the molecule changes its molecular structure due to exposure to light. That is, their atomic arrangements are altered. These molecules exhibit two distinct states: a thermally stable {\em trans}-state, often denoted as "E" and a less stable {\em cis} configuration, known as the "Z" state. Fig. \ref{figazo} shows the conversion of two states on exposure to light. When light is within the broad absorption band of the trans-azo state, photochemical isomerization occurs. Subsequently, the {\em cis} configuration typically returns to the thermally stable {\em trans}-state. The time it takes to do so depends on the specific substitution pattern of the azobenzene molecule and local conditions. Additionally, irradiating the {\em cis} form with light within its distinct absorption band can trigger the photochemical cis-to-trans isomerization Fig. \ref{figazo}a. This straightforward photochemistry underpins the diverse photoswitching capabilities that lead to different molecular gates such as SPARK that are used in many applications \cite{chambers2008light},\cite{xue2021fluorescent}.\par
 
 \paragraph{SPARK} channels incorporate azobenzene molecules to control the flow of potassium (K$^+$) ions through the channel. This gate comprises three components: a genetically engineered K$^+$ channel protein, a pore-blocking element, and a photoisomerizable azobenzene molecule. When exposed to long-wavelength light (around $500$nm), the azobenzene segment adopts its elongated {\em trans} configuration, enabling the blocker to inhibit ion flow. Conversely, upon exposure to short-wavelength light (around $380$nm), the azobenzene adopts its shorter {\em cis} configuration, retracting the blocker and permitting ion conduction through the channel Fig. \ref{figazo}b. \cite{yager2006novel},\cite{fortin2011engineering}.\par
 \rev{ Though SPARK channels are designed to be photosensitive, thermal energy can also induce transitions. However, the effect is much slower than the light-induced transitions \cite{yager2006novel}. 
 The work in \cite{liu2020alkyl} and \cite{tamaoki2013photocontrol} highlights the precise and rapid nature of light-induced switching in azobenzene-based systems, such as SPARK channels in our case, noting that thermal relaxation from the {\em cis} to {\em trans} state is slow and can take hours under room temperature conditions. The transition between the {\em trans} and {\em cis} states of azobenzene occurs on the timescale of picoseconds to nanoseconds, which allows for almost instantaneous changes in the channel state upon light activation \cite{lin2023investigation},\cite{koch2021cyclic},\cite{tamaoki2013photocontrol}.}\par
Fig. \ref{figazo}c shows states represented by a compound label consisting of state property and state number. We have designated the {\em trans} state as {\em C1} and the {\em cis} state as {\em O2} to remain consistent with our previous work. A solid, thick arrow depicts an input-dependent transition, the light level in this case. State transitions are defined by transition rates that determine the probability of a state change occurring in an infinitesimally small amount of time. 
 \textcolor{black}{\paragraph{Measurement of SPARK Channel States} The states are measured through the detection of ion flow, which serves as an observable output of the system. To quantify these transitions, electrophysiological techniques such as patch-clamp recordings \cite{sigworth2007patch}, \cite{kim2011electroanalytical} or fluorescence-based assays can be employed. Patch-clamp techniques provide a direct measurement of ionic currents across the SPARK channel by recording changes in membrane potential or current fluctuations. Alternatively, fluorescence-based methods \cite{voldvrich2024fluorescence} utilizing voltage-sensitive dyes or Förster resonance energy transfer (FRET)-based reporters can be used to monitor the conformational changes associated with SPARK activation.}

\textcolor{black}{In our model, the state of the SPARK channel is inferred based on these established experimental methods, with transitions between states being governed by the probabilistic framework outlined in the master equation.}

\subsection{Mathematical model}
Following the principles of mass action kinetics, the chemical kinetics of the receptor are governed by a differential equation known as the master equation, as described in \cite{swain1984handbook}, \cite{channelrhodopsin2018}. This equation incorporates two key parameters: the rate matrix $Q$ and the probability of the receptor being in state $i$ at time $t$. The rate matrix $Q=[q_{ij}]$ represents the $k\times k$ array of per capita transition rates, where the diagonal elements are constrained such that the sum of each row is zero, and each $q_{ij}$ denotes the instantaneous transition rate from state $i$ to state $j$. When visualizing $Q$ as a graph, the states correspond to $k$, and a directed edge is drawn from vertex $i$ to $j$ if $q_{ij}>0$.

\textcolor{black}{To model the effect of light on the transition dynamics, we assume that the transition rates are linearly dependent on the light intensity. Our assumption is justified based on the experimental findings in \cite{ishizuka2006kinetic}, demonstrating that the photocurrent is linearly dependent on light intensity. Given that the SPARK channel's gating mechanism is directly controlled by photoisomerization of azobenzene, we draw parallels between the observed linear photocurrent response and the probability of state transitions in SPARK.
}

\begin{figure}[h!]
\centering
\includegraphics[width=\columnwidth]{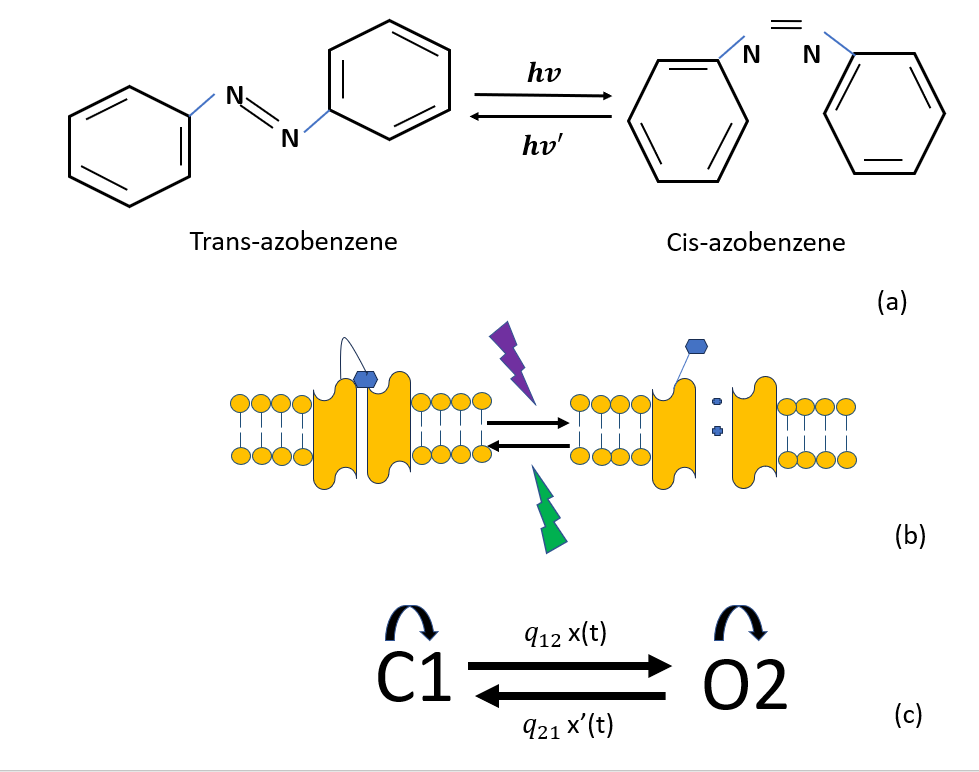}
\caption[\hspace{0.2cm}]{(a) Azobenzene structure in 'trans' and 'cis' state. Formational change occurs when exposed to the light (b) SPARK is formed when shaker $K^+$ protein and blocker are attached to azobenzene structure. Figure adapted from \cite{xue2021fluorescent},\cite{chambers2008light}. (c) Two-state model}
\label{figazo}
\end{figure}

As mentioned earlier, both transition states are affected by variations in light intensity we denote $x(t)$ as the light intensity required to transition from state {\em C1} to {\em O2}, and $x'(t)$ as the light intensity necessary for the transition from state {\em O2} back to {\em C1}. For instance, $q_{12}x(t)$ signifies the transition rate from state {\em C1} to {\em O2}, which is responsive to changes in $x(t)$. Similarly, $q_{21}$ represents the transition rate from {\em O2} to {\em C1}, which depends on external input $x'(t)$ Fig. \ref{figazo}c.
      
For a receptor comprising $k$ distinct states, there is a k-dimensional vector denoted as $p(t)$, as described in  \cite{channelrhodopsin2018}
\begin{align}
   \label{eqn:state probabilities}
     p(t) &= [p_1(t),p_2(t),. . ., p_k(t)] ,
 \end{align}
 In this equation, $p_i(t)$ signifies the probability of the receptor being in state $i$ at time $t$.

Using the rate matrix $Q$ and the vector $p(t)$, the master equation can be expressed as:
\begin{align}
   \label{eqn:master equation}
     \frac{dp(t)}{dt} &= p(t)Q ,
 \end{align}
 
\begin{table}[ht!]
    \centering
    \begin{tabular}{ | m{3.2cm} | m{3.2cm}| m{1cm} | } 
    \hline
     Parameter & Rate values & Units \\
     \hline
     $q_{12}$x(t) & $0.2x(t)$ & $s^{-1}$\\
     \hline 
     $q_{21}=q_{dark}+(q_{light})x'(t)$ & $0.0037+0.08(x'(t))$  &      $s^{-1}$\\
    \hline
\end{tabular}\\
    \caption{Rate Parameters for SPARK from \cite{chambers2008light}}
    
    \label{tab:Table I}
\end{table}
\rev{Table I shows the rate parameter values from the literature \cite{chambers2008light}.}
The second row of Table I shows two values for the transition from {\em cis} to {\em trans} state that is $q_{dark}$ and $q_{light}$. These two values are the transition rate from open to closed in different environments. $q_{dark}$ is the rate when there is no light whereas $q_{light}$ is the constant used to determine the value of $q_{21}$ when there is light.\par
\rev{While the transition rates are measured in units of {\em transitions per second (s$^{-1}$)}, the intensity is measured in irradiance or W/m$^2$. Thus, irradiance-dependent rate constants (e.g., $q_{12}$) are given in units of m$^2$ W$^{-1}$ s$^{-1}$.
}
 
\rev{ Analyzing communication channels in continuous time using the continuous-time master equation poses significant mathematical challenges. As an alternative, a discrete-time, discrete-state representation of the master equation was introduced in \cite{channelrhodopsin2018}; we review this model below.} By using a discrete-time approximation to the master equation, we have 
\begin{align}
   \label{eqn:discrete time}
     \frac{dp}{dt} &= \frac{(p(t+\Delta t))-p(t)}{\Delta t} + o(\Delta t) = p(t)Q,\  as\  \Delta t\rightarrow 0 \\
     p(t+\Delta t) &= \Delta t p(t)Q(t) +p(t) + o(\Delta t)\\
                   &= \Delta t p(t)Q(t) +p(t)I + o(\Delta t)\\
                   &= p(t)(I + \Delta t Q)+o(\Delta t)\   as\   \Delta t\rightarrow 0
\end{align}
where $o(\Delta t)$ represents a term that is negligible as $\Delta t \rightarrow 0$.
For a discrete-time model, we introduce the approximation
 \begin{align}
   \label{eqn:discrete time2}
     p_i = p(i\Delta t)+o(\Delta t),\ as\ \Delta t \rightarrow 0\,,\\
     p_{i+1}=p_i(I+\Delta t Q)\,.
\end{align}


Based on \cite{channelrhodopsin2018}, we describe the probability transition matrix for discrete-time Markov chains using  as follows:
 \begin{align}
   \label{eqn:transition matrix}
   P &= I+\Delta t Q\,,
 \end{align} 


 %
 \rev{The state transition matrix for SPARK is} \\\\\\
\begin{equation}
 \label{eqn:transition matrix values}
  P = \begin{bmatrix}
1 & 0 \\
0 & 1 

\end{bmatrix}
+\Delta t
\begin{bmatrix}
 -q_{12}x(t)& q_{12}x(t) \\
  q_{21}x'(t) & -q_{21}x'(t)
\end{bmatrix}
\end{equation}\\

 We will briefly explain the variables used in the equation:
 \begin{itemize}
     \item $Q$ is the rate matrix defined in the previous section
     \item $I$ is the identity matrix of the appropriate size. 
     \item $\Delta t$ is a time step introduced to study the properties of the receptor as a discrete-time, discrete-state channel influenced by a continuous time channel. The time step $\Delta t$ should be small enough so that $P$ meets the conditions of the Markov chain transition matrix (row-stochastic and nonnegative).
 \end{itemize}

%
The rate matrix $Q$ (on the right side of $\Delta t$) shows the possible transition rates. Here , we set $-q_{12}x(t)$ and $-q_{21}$  to make row sums equal to zero. 

\section{Communication model}
We aim to present an innovative communication system that utilizes the biological receptor SPARK. We will initially analyze its functionality using a single receptor and expand our study to include multiple receptors. The following section will elaborate on viewing SPARK as the receiver in a light-based communication system.

\subsection{Communication system components}
The components of the communication model in Fig. \ref{fig1} are described as follows.
\begin{itemize}
    \item {\em Transmitter.} The transmitter comprises three key components: an encoder, a modulator, and a light source. Initially, the input bit stream is encoded, resulting in an encoded bit stream. The data is then modulated and directed to the light source via the optical channel for transmission. For our configuration, we adhere to the arrangement described in \cite{khan2017visible}, which presumes the utilization of two laser sources, one (380nm) to open the SPARK channel and the other (500nm) to close the channel. 
    \item {\em Channel.}
     The channel is an environment which can facilitate the flow of light. It can be free space or any other optical channel, such as optical fibre.
     \item {\em Receiver.}
     Typically, visible light communication (VLC) signals are received using a photodiode, as described in reference \cite{10.1145/3576781.3608721}. 
     Any current-measuring technique mentioned in Section II.B can be employed to capture the current. The transmitted message can be detected using a demodulator and decoder in this setup.

\end{itemize}
\subsection{Channel Model}
We will now discuss the channel input and output state relationship of receptors. The matrix $P$ is a non-negative and row-stochastic matrix that meets the conditions of a Markov chain transition probability matrix as long as observation time $\Delta t$ is small enough (\ref{eqn:transition matrix values}). We will use the same notations as mentioned in \cite{channelrhodopsin2018}.\\
\begin{itemize}
    \item {\em Input:} \rev{
    We have assigned the value of the light intensity  $x(t)$ to represent bit '1' and of $x'(t)$ (in units corresponding to the rate in Table \ref{tab:Table I}) to represent bit '0'.
    In subsequent sections, we will vary the input $x'(t)$ within a specific range and observe its effects on the properties of the channel.}
    \item {\em States}: The output in this case is the state of the receptor represented by $s_{i}$. As in Fig. \ref{figazo}a, the receptor may be in one of two states, the channel state $s_{i}$ is represented by a number and is also discretized. Hence, the output vector would be $\textbf{s}=[s_{1},s_{2},...,s_{n}]$.
    \item {\em Input-State response}: The relationship between channel input and output can be described as a PMF
     \begin{align}
        \label{eqn:psgivenx}
         p(\textbf{s}|{x}) & = \prod_{i=1}^{n}p(s_{i}|x_{i},s_{i-1})
     \end{align}
    where $p(s_{i}|x_{i},s_{i-1})$ is given by transition matrix P. The transition from state $s_{i-1}$ to $s_{i} $ is considered \emph{insensitive} if for all $x_{i}\in \chi$, $p(s_{i}|x_{i},s_{i-1})= p(s_{i}|s_{i-1})$. In our case, if the transition from state $C \rightarrow O$ is sensitive to input then the \emph{self transition} $C\rightarrow C$ is also sensitive to input. 
\end{itemize}

    
\subsection{Modulation technique}

\begin{figure}[t!]
\centering
\includegraphics[width=\columnwidth]{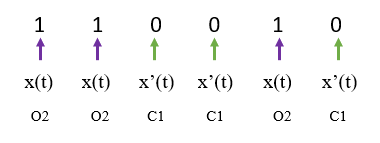}
\caption[\hspace{0.2cm}]{Illustration of modulation technique. A bit sequence can be sent using light pulses $p_1$ and $p_0$.}
\label{figcom}
\end{figure}

The modulation technique used is similar to that of frequency shift keying combined with on-off keying. The encoding of the subsequent bit depends on the present one. 
The different light pulses illuminated for a specified amount of time. \textcolor{black}{ Each bit is transmitted over a fixed interval 
$T$, during which the corresponding light source ($380$nm for bit '1' or $500$ nm for bit '0') remains active}. Fig. \ref{figcom} shows the sequence of bits having repeated patterns of $1$'s and $0$'s; when there is a bit $1$, the state changes from {\em C1} to {\em O2} and vice versa. If the current bit is the same as that of the preceding one, the same state is assumed, and consequently, the same pulse is used to maintain the channel in its current state.  This property differs from channelrhodopsin (ChR2), where one has to wait for the complete cycle to send another bit.\par

 Generally, it will be convenient to represent  $\vec{b}$ as a vector of pulses. For example, a light of frequency $ ~380$nm is generated to send bit $1$. The detector observes the receptor and records its status in $n$ regular intervals of $\Delta t$. If there is a current or a state transition between $C1$ and $O2$, it is interpreted as bit 1; otherwise, it is bit $0$. The number of observing intervals $n$ would be considered as observing states that could be $3,4,5,...$ and $n\Delta t$ constitute the total time interval $T$.
Suppose the transmitter releases $m$ bits. The input pulses for these bits can be represented by $\Vec{x}=[x_1(t),x_2(t),...x_m(t)]$ or $\Vec{x'}=[x_1'(t),x_2'(t),...x_m'(t)]$ depending on bits; for example, if $b_i = 1$ and $b_{i+1} = 0$ then the release time of $x_i$ is $iT$ and the release time of $x'_{i+1}$ is $(i+1)T$.



\begin{figure}[ht!]
\centering
\includegraphics[width=\columnwidth]{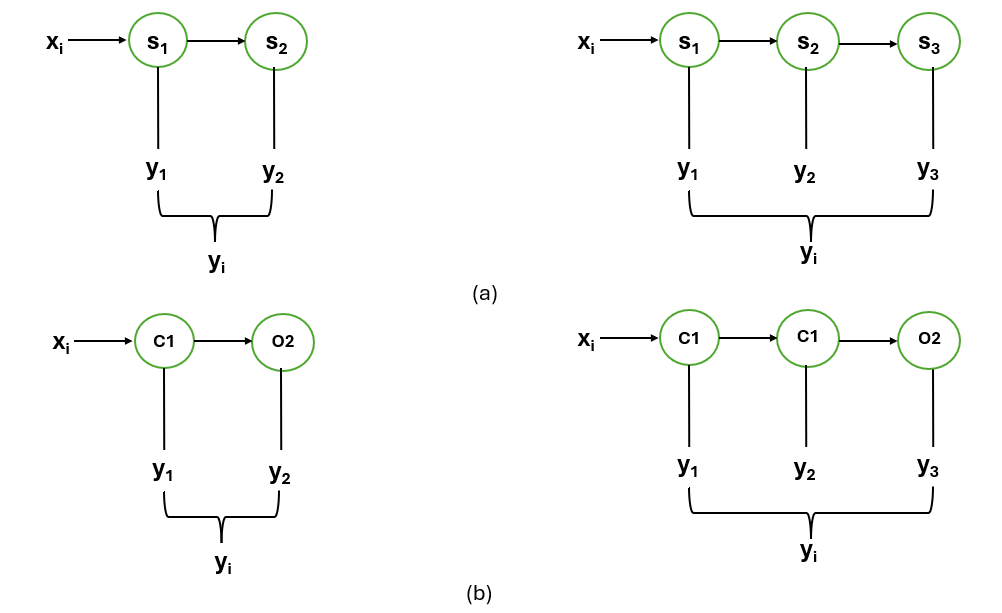}

\caption{State combination example:(a) Input $x_{i}(t)$ causes state changes which is measured in interval of time $\Delta t$, output $y_i$ is detected after observing each output after sampled interval, (b) is an example of state combinations of SPARK channel}
\label{Fig4}
\end{figure}
\subsection{Receiver and Detection}

For the purpose of detection, we aim to identify bit $i$ as a vector  $\Vec{y_i}$. Thus, the vector $\Vec{{y}_i} = {y_1, y_2, y_3, ..., y_n}$ represents the outputs of each observed state within a discrete-time interval $\Delta t$.
The decision is made by assessing the aposteriori probabilities computed for each state combination, as Table II outlines for $n = 2$. Fig. \ref{Fig4}a and b show the detection process for combinations involving two and three states, respectively. For instance, in the scenario where $n = 2$, implying that the receptor is observed for $2\Delta t$ times intervals, there are $2^n$ state combinations, two of them are detected by the presence of current. The decision rule is established by evaluating the posterior probability for each state combination:
 
\begin{align}
 \label{eqn:aposteriori}
     p(x|\textbf{y}) &= \frac{p(\textbf{y}|{x})p(\textbf{x})}{\sum_x p(\textbf{y}|\textbf{x})p(\textbf{x}) }\, ,\\\\
                              &= \frac{\sum_{i=1}^{n}p(y_1,...,y_n|s_i)p(s_i|x)p(x)}{\sum_x \sum_{i=1}^{n}p(y_1,...,y_n|s_i)p(s_i|x)p(x)}\, , 
 \end{align}
 where
\begin{align}
     \label{eqn:pygivens}
     p(y|s) &= 
     \left\{
        \begin{array}{cl}
             1, & y=1,\,s=O; \\
             1, & y=0,\,s=C; \\
             0, & otherwise
        \end{array}
    \right.
\end{align}
and $p(s|x)$ can be determined from (\ref{eqn:psgivenx}) using transition matrix $P$ from (\ref{eqn:transition matrix}).\\
If
 \begin{align}
     \label{eqn:comparison}
     p({x=0}|\textbf{y}) > p({x=1}|\textbf{y})\rightarrow 0\, ,\\
     p({x=0}|\textbf{y}) < p({x=1}|\textbf{y})\rightarrow 1\,.
\end{align}
The probabilities and observable state combinations calculated from equations (\ref{eqn:aposteriori})-(\ref{eqn:pygivens}) are displayed in Table II.

As an example of our detection mechanism, suppose the detector observes {\em no current} in the first interval $\Delta t$ but detects {\em current} in the second interval, so the $\vec{y} = [0,1]$ as an output $[y_1,y_2]$. Table II shows that this output corresponds to (C1, O2). The transition rates from Table \ref{tab:Table I} are used to derive the probabilities. Based on the decision rule, (\ref{eqn:comparison}-9) the detector will decide in favour of bit '1' assuming that by turning on the light pulse $p_1$, the transmitter has sent bit '1'.\par

Consider another example of sending bit '0' and the detector observes {\em an ion current} in the first interval and {\em no current} in the other interval so that the corresponding output will be considered as $\vec{y}=[1,0]$. According to rows (3) in Table II, $p(x=0|y) > p(x=1|y)$, so the receiver will decide in favour of '0'. However, in the other two cases, there is a minute difference in the probabilities. If the detector observes {\em no current} for two discrete time intervals, then it indicates (C1, C1) that are $\vec{y}=[0,0]$. The message will be considered bit '0' for $p(x=0|y) > p(x=1|y)$. The probabilities can be determined for any number of states. However, we have considered sequences for two and three states, which are observed over intervals of $\Delta t$ for a total time of $T$.

\begin{table}[!h]
    \centering
    \begin{tabularx}{0.5\textwidth} { 
  | >{\raggedright\arraybackslash}X 
  | >{\centering\arraybackslash}X 
  | >{\raggedleft\arraybackslash}X | }
  \hline
  State Combinations & $p(x|y),x=100$ W/m$^2$ & $p(x|y),x'=100$ W/m$^2$ \\
  \hline
  C1-C1 & $0.40$ & $0.6$\\
  \hline 
  C1-O2     & $1.0$        &        $0$ \\
  \hline
  O2-C1     & $0$        &        $1$\\
  \hline
  O2-O2     & $0.54$        &        $0.46$\\
  \hline
 
\end{tabularx}
    \caption{A posteriori probabilities of states when $n=2$.The transition rates $q_{12}x(t), q_{21}x'(t)$  are used to determine these probabilities}
    
    \label{tab:Table II}
\end{table}

 Data rates can be determined by
\begin{align}
    \label{eqn:DataRate}
    R = \frac{1}{T}\, . 
\end{align}
in bits per second. This simplified rate is sufficient to demonstrate the feasibility of the scheme. We leave the more sophisticated design to future work.

\subsection{Multiple Receptors}

\rev{This section will expand our model to include multiple SPARK receptors. Each receptor operates independently of the others. The opening or closing of one channel does not have an impact on the others. 
The purpose of taking multiple receptors in account is to reduce waiting time as there is a high chance of opening at least one receptor in response to input, thus increasing data rates.}
 For example, when there are two receptors, the waiting time to detect a bit (ion current) is shorter than that of only one receptor. In order to find out {\em transition probability matrix}, we have determined possible combinations of two receptors in three states, as shown in Table III. The first row of the table contains the number of states $C1$ and $O2$, whereas the other rows indicate the number of receptors in each state. For example, the second row shows that there is one receptor in each state. Similarly, in the third row, there are two receptors in the state $O2$. 

\textcolor{black}{While some classes of cooperatively-gating receptors have been found \cite{dixon2022mechanisms}, it is common and very widely assumed in the literature that different ion channels are gated independently of each other \cite{anderson2015stochastic}. This assumption is extensively used in the literature (e.g. \cite{thomas2016shannon,moffett2022permissive}), so we take the independent operation of SPARK channels as a reasonable assumption to use in this paper.}
 
 The state table helps to determine transitions of receptors from one state to another, leading to the probability matrix. In order to find the total number of possible combinations, the {\em star and bar method} from \cite{feller1991introduction} can be used. In this method, if there are $[N]$ items and $k$ bins, $k-1$ bars or separators are needed in order to get $N$ items into $k$ bins. There are $(N+k-1)$! permutations of ordering $n$ items. The permutation of $n$ items does not matter, and the permutation of $(k-1)$ separators does not matter. \textcolor{black}{Applying this method, the total number of possible combinations is given by}:\\
\begin{align}
    \label{eqn:combination}
    \binom{N+k-1}{k-1} = \frac{(N+k-1)!}{N!(k-1)!}
\end{align}
\begin{figure}[ht!]
\centering
\includegraphics[width=\columnwidth]{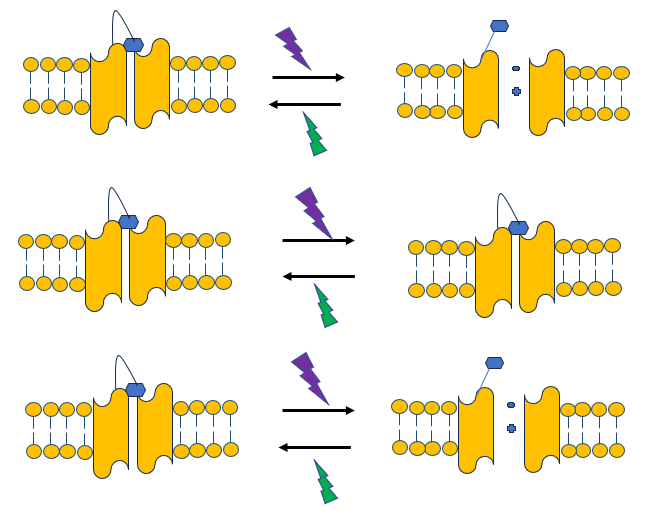}
\caption{ Three SPARK receptors for illustration. The first and third ones are open, and the second one is closed. It shows that a number of SPARK channels work independently}
\label{fig2chR2}
\end{figure}

From (\ref{eqn:combination}), if we have $N=2$ SPARK receptors and $k=2$ states, then there are three combinations listed in Table III.\\

\begin{table}[ht!]
    \centering
    \begin{tabularx}{0.4\textwidth} { 
  | >{\raggedright\arraybackslash}X 
  | >{\centering\arraybackslash}X 
  | >{\raggedleft\arraybackslash}X| }
  \hline
  C1 & O2  \\
  \hline
  $2$ & $0$ \\
  \hline 
  $1$  & $1$  \\
  \hline
  $0$   & $2$ \\
  \hline
  
\end{tabularx}\\
    \caption{ combinations of two receptors in three states}
    \label{tab:Table III}
\end{table}
Following the process, the state transition diagram can be determined by considering the possibilities of transitions from one state combination to another. 
 
 We have two receptors in combination state $C1$ in Table III, and there is a chance that one of the receptors changes its state from $C1$ to $O2$, which is mentioned in the second row of Table III.
\begin{figure}[h!]
\centering
\includegraphics[width=0.9\columnwidth]{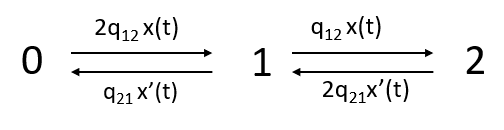}
\caption[\hspace{0.2cm}]{Directed graph between states when there are two receptors.}
\label{figtransition}
\end{figure}
 In the same way, both receptors can change their states to $O2$. The state diagram is shown in \rev{Fig. \ref{figtransition}.}
The transition probabilities are determined using the rate matrix from Table \ref{tab:Table I}. For instance, the transition probability from $0$ to $1$ represents that any other receptor can change its state from $C1$ to $O2$ with probability \rev{$2q_{12}x(t)$  whereas $q_{12}x(t)$} is the probability of transition when both receptor move to the state.

The probability transition $P_c$ matrix can be determined from the state diagram showing all possible transitions from one combination state to another. 



\begin{equation}
 \label{eqn:state matrix values}
  P_c = \begin{bmatrix}
1-2q_{12}x(t) & 2q_{12}x(t) & 0 \\
q_{21}x'(t) & 1- (q_{21}x'(t) + q_{12}x(t)) & q_{12}x(t) \\
   0   & 2q_{21}x'(t) & 1 - 2q_{21}x'(t) \\
\end{bmatrix}
\end{equation}\\
The generalized process of generating a probability transition matrix for any number of multiple receptors can be described as follows:
\begin{itemize}
    \item For $N$ number of molecules with let's say two discrete states $C1$, $O2$, calculate number of combinations using (\ref{eqn:combination})
    \item Make a state table with two columns showing the number of molecules in each state. Fill in each row with a receptor,  as illustrated in Table III, showing different numbers of receptors in different states. 
    \item Draw a state diagram by connecting all possible state combinations in the form of a directed graph where each edge corresponds to the edge of the rows and arrows indicate transition probabilities.
    \item Create a probability transition matrix based on the rate matrix with rows and columns equal to the number of state combinations calculated from (\ref{eqn:combination}).
    
\end{itemize}
\label{eqn:intensityandphoton}

\section{Results and Discussion}

\rev{ \color{black}{Monte Carlo simulations were conducted to estimate the probability of error and data rates under various system configurations, ensuring statistical robustness and mitigating stochastic variations. Each iteration generated a random binary sequence of 100 bits, representing the transmitted signal, with state transitions modeled using a Markovian framework described in Section II.C. The transition probabilities between these states are computed as per (\ref{eqn:transition matrix}), and the state evolution follows a probabilistic random walk process. The total simulation time is discretized into small steps of $\Delta t$, with multiple observation states $n$ such as $2$, $8$, $10$, $20$, $50$ used to examine different temporal resolutions. 
%
%
A default time step of
$\Delta t = 2$ ms is chosen to capture the dynamics of the system, however, $\Delta t$ is also varied to determine any effects of discretization. 
%
%
and the probability of error is computed by comparing received and transmitted bits. Data rates are inferred from the effective time duration required per bit transmission. The results are averaged over all simulation trials, and plots are used to visualize the probability of error as a function of total time, highlighting system behaviour across different parameter configurations.}}\par
\rev{We define a minimum acceptable error rate of $10^{-2}$, which is appropriate for this communication system based on the expected noise levels in optical and biological environments \cite{nakano2013molecular},\cite{proakis2008digital}. This error rate is typical in systems that employ error correction techniques, such as forward error correction (FEC), to ensure reliable communication. The system parameters, including $\Delta t$, $T$, and the number of receptors $n$, are chosen to see the error performance.}

\subsection{Effect of controlled light intensity}
\textcolor{black}{The results in this section demonstrate that controlling light intensity plays a crucial role in influencing data rates and error probability in the SPARK channel. Specifically, the probability of error has been estimated as a function of the total time interval $T$, using a fixed observing interval $\Delta t$ and a defined number of observations $n$.}\par

\begin{figure}[hbt!]
\centering
\includegraphics[width=\columnwidth]{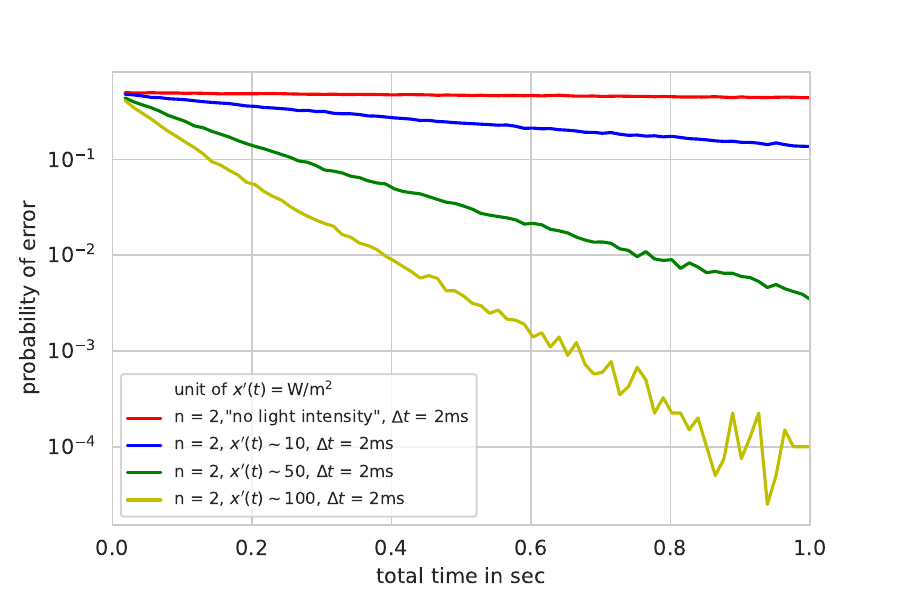}
\caption{Probability of error vs. total time: Impact of light intensity on the reverse transition from cis to trans state.}
\label{fig_light_intensity_var}
\end{figure}

\begin{figure}[h!]
\centering
\includegraphics[width=\columnwidth]{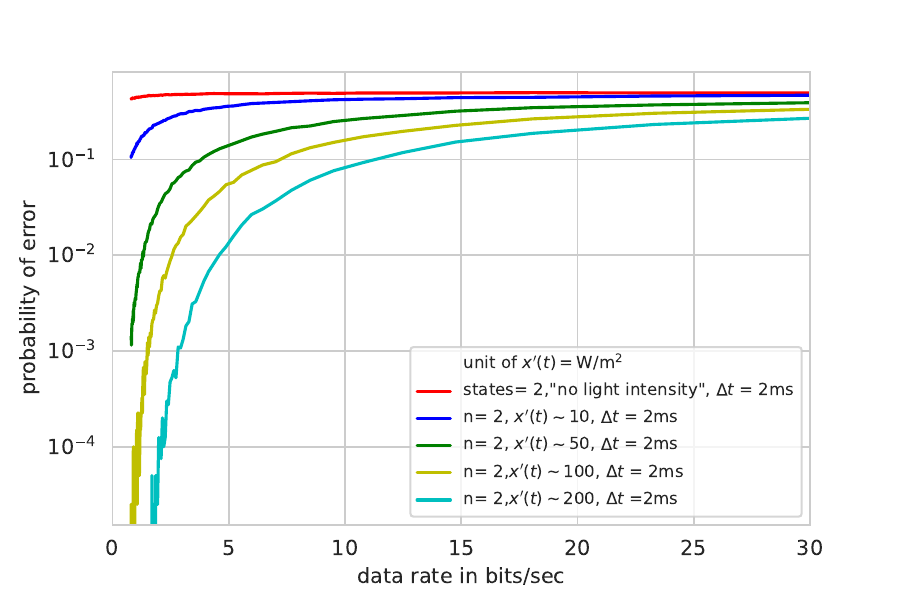}
\caption{The figure shows the trade off between data rate and probability of error at different light intensities.}
\label{fig_datarate_intensity}
\end{figure}

By adjusting the intensity of light, we actively controlled the transition of SPARK molecules back to their {\em trans} state, thereby demonstrating the impact of light control on communication performance. As shown in Fig. \ref{fig_light_intensity_var}, the probability of error remains significantly high when the transition back to the {\em trans} state is not regulated. Similarly, Fig. \ref{fig_datarate_intensity} illustrates that in the absence of controlled light intensity, there is substantial variability in data rate with respect to the probability of error, further underscoring the importance of light modulation in optimizing the system's performance.

\subsection{Single Receptor}
In this subsection, we present our results in terms of the probability of error as a function of input, $\Delta t$ and total time $T$. We will explicitly express the effects of the number of states on data rates. Our system is designed to observe the receptor over time intervals $2\Delta t$ ($n=2$), $3\Delta t$ ($n=3$), $4\Delta t$ ($n=4$), $8\Delta t$ ($n=8$), $10\Delta t$ ($n=10$), $20\Delta t$ ($n=20$), $50\Delta t$ ($n=50$)  time intervals with specific values of $\Delta t$.
\begin{figure}[hbt!]
\centering
\includegraphics[width=\columnwidth]{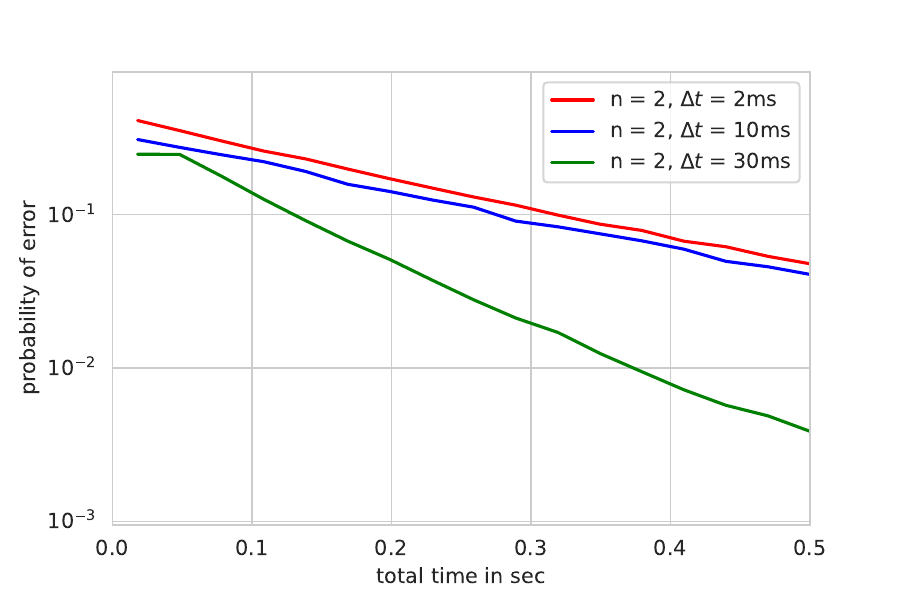}
\caption{Probability of error vs total time where $\Delta t$ is variable for observation states $n=2$.}
\label{figtotaltimevspbspark1}
\end{figure}

\begin{figure}[hbt!]
\centering
\includegraphics[width=0.5\textwidth]{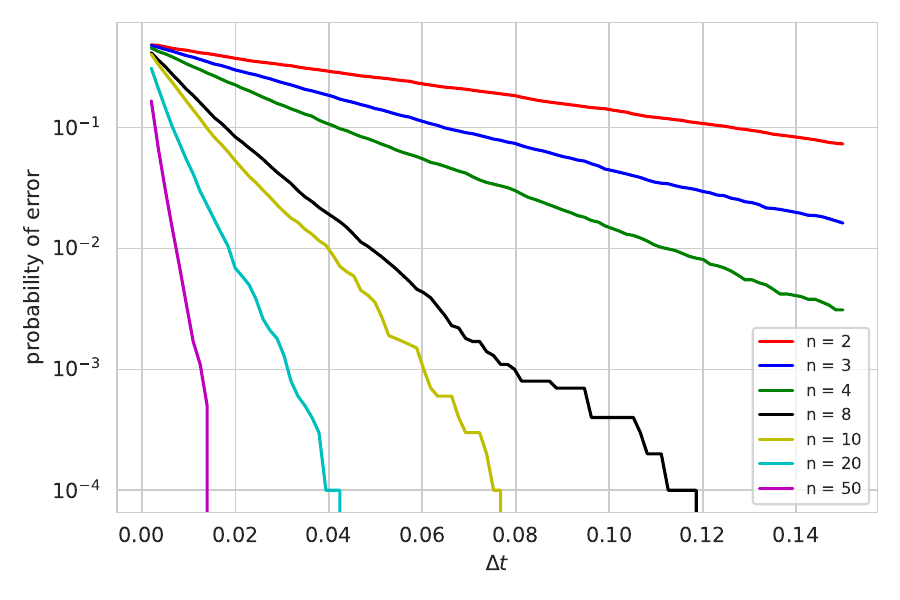}
\caption{Probability of error decreases linearly with increasing $\Delta t$ in log scale for different values of states.}
\label{figdifferentn}
\end{figure}

In Fig. \ref{figtotaltimevspbspark1}, the focus is on understanding how variations in total time and time steps affect the likelihood of errors in a system with a specific state $n = 2$. \rev{The graph reveals a notable shift in the probability of error by examining the time steps of $\Delta t=2,10,30$ ms. We can see that the probability of error meets the requirement of our minimum bit error rate at $t = 30$ ms.} Particularly, there's a striking difference in error probability as the time step $\Delta t$ increases from $2$ms to $30$ms. This finding suggests that the timing at which observations are made significantly impacts the system's error rate.
Moreover, the graph indicates an overarching trend: as the total time interval $T$ increases, the probability of error decreases. This observation aligns with an intuitive understanding that longer observation times reduce the chances of errors. Observation states, or how often the detector observes the receptor before making a decision, also affect the error rate. Fig. \ref{figdifferentn} compares the error probability for different numbers of observation states. Though it decreases the probability of error for a small total time $T$, it will increase the system's complexity.



\begin{figure}[ht!]
\centering
\includegraphics[width=\columnwidth]{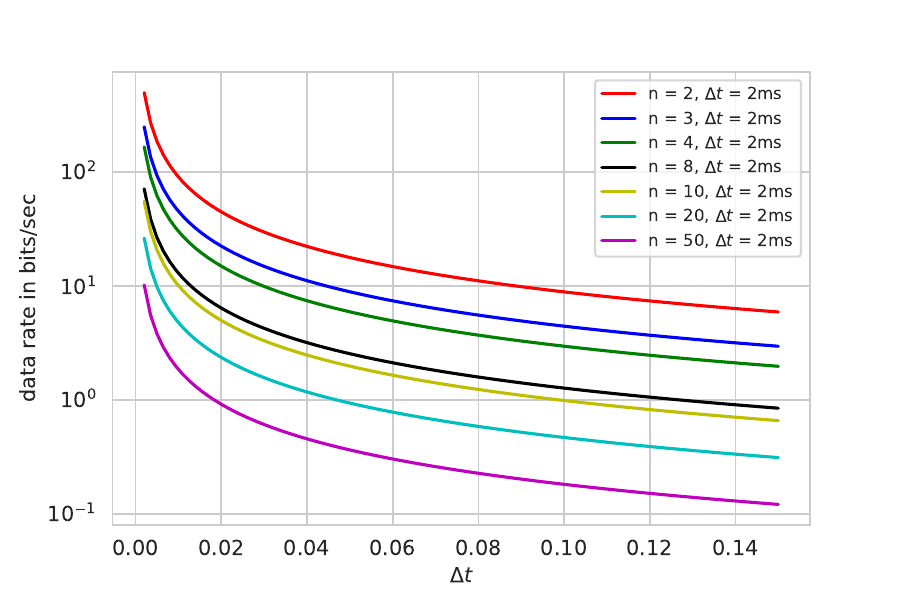}
\caption{Data rates become lower as the number of observation states increases with $\Delta t$. }
\label{figdataratedifferentn}
\end{figure}

\begin{figure}[ht!]
\centering
\includegraphics[width=\columnwidth]{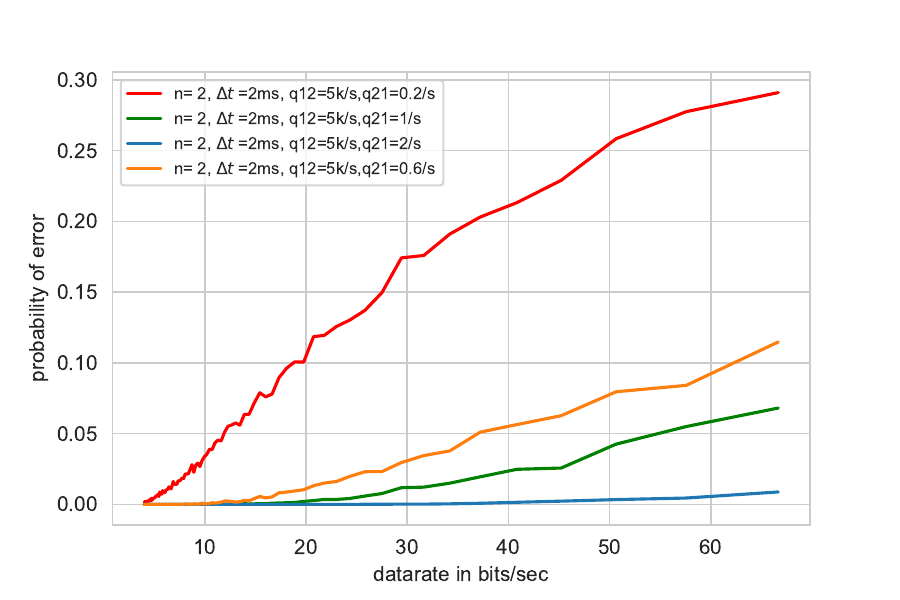}
\caption{Probability of error vs data rate for single SPARK molecule when chemical rates of transitions were different. The rate of transition affects the information rate. The higher the chemical rate of transition, the higher the data rate.}
\label{figdifftransitionrates}
\end{figure}
 In Fig. \ref{figdifferentn}, we compare the probability of error for different observing states $n=2, 3, 4, 8, 10, 20, 50$ with the time steps $\Delta t = 2$ms. The results show that the probability of error for states $n=2$ is initially higher $(12\%)$ than that of $n=3$.\par


\subsection{Data Rates} 





\begin{figure}[ht!]
\centering
\includegraphics[width=\columnwidth]{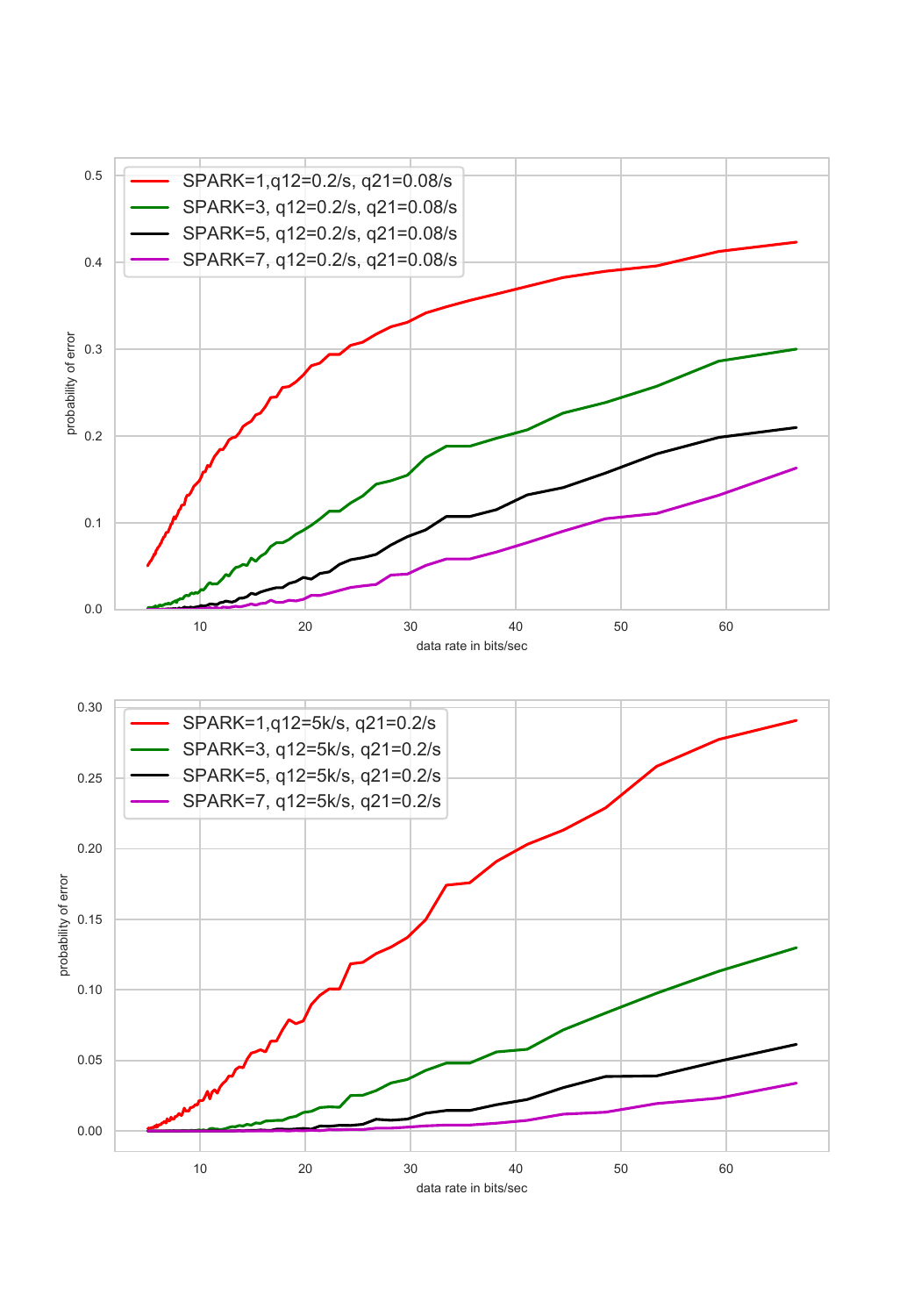}
\caption{Probability of error vs datarate for multiple SPARK channels when the chemical rate of transition is different. }
\label{figmultiplespark1}
\end{figure}
\textcolor{black}{The data rate of our system is determined by the switching dynamics of the SPARK channel, which responds to light stimuli by opening or closing to allow ion flow.  Experimental studies on ion channels, \cite{moffett2022permissive}, have demonstrated that ion current transitions occur without measurable delay upon channel opening or closing. Therefore, our data rate analysis focuses on the temporal dynamics of the channel gating rather than the diffusion process of the ions.}
based on our definition of data rate from (\ref{eqn:DataRate}). We have analyzed that there is a trade-off between data rates and the probability of error. Fig. \ref{figdataratedifferentn} shows estimated data rates with respect to $\Delta t$ for $n=2, 3, 4, 8, 10, 20, 50$. It shows that the data rate for higher states is lower than that of  $n=2$ due to an increase in "observation time" with the number of states. As the observation time increases, the total time $T$ also increases, leading to a decrease in the data rate. Though the system has a low data rate, it experiences a low error probability with an increase of $\Delta t$ as explained in Fig. \ref{figdifferentn}.\par 
 It can be seen that the probability of error increases with the data rate for the same $\Delta t$, as shown in Fig. \ref{figdifftransitionrates}. We have added another parameter in our analysis: the transition rate from one state to another, which depends on the chemical rate of the biomolecule. There are ongoing efforts to make the photo switch faster \cite{szymanski2019}. So, we have done simulations with different transition rates. In Fig. \ref{figdifftransitionrates}, we increase the rate $q_{12}$ (rate from state $1$ to state $2$) up to $5$ k/s and simulate for different values of $q_{21}$ (the rate from state $2$ to $1$). With $n=2$, with $\Delta t= 2$ms and $q_{12}$ fixed at $5$ k/s, the probability of error is only $5\%$ for data rate of $20$ bits/s when $q_{21}=0.6$/s. However, for similar data rate but for less $q_{21} = 0.2$/s, the probability of error increases to almost $20\%$.

\begin{figure}[h!]
\centering
\includegraphics[width=\columnwidth]{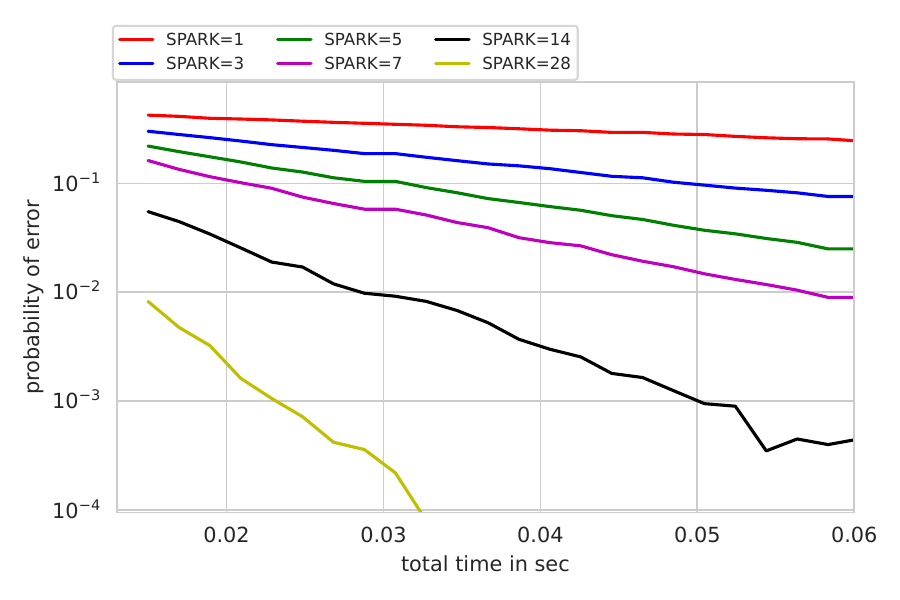}
\caption{Probability of error vs total time for different numbers of SPARK. There is a significant decrease in the probability of error as the number of molecules increases.}
\label{figmultiplespark2}
\end{figure}

\begin{figure}[h!]
\centering
\includegraphics[width=\columnwidth]{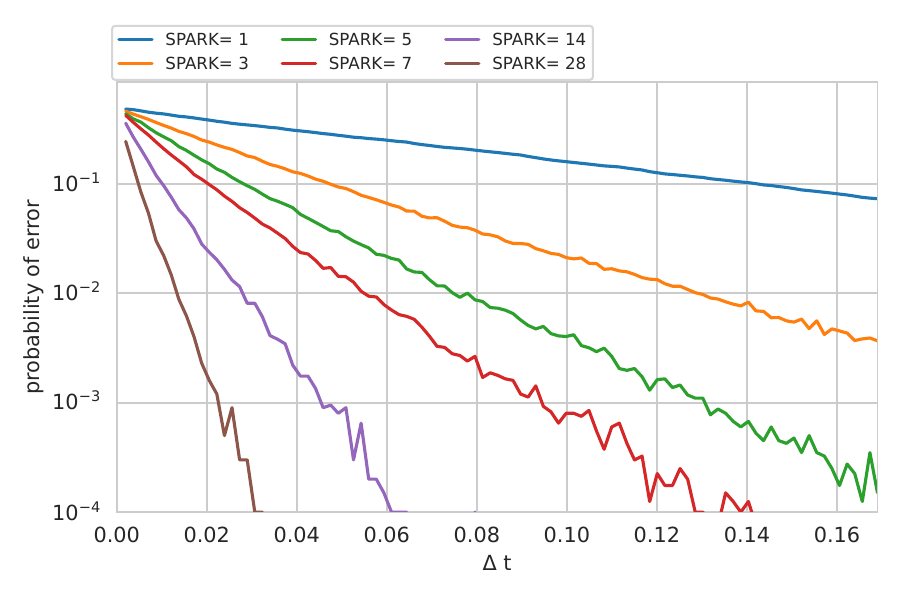}
\caption{Comparison of multiple SPARK: Probability of error vs $\Delta t$. The probability of error is reduced as the number of SPARK increases.}
\label{figmultiplespark3}
\end{figure}

\subsection{Multiple Receptors: Improvement of Data Rates}
In this section, we will show that other than faster chemical rates, it is possible to obtain higher data rates by using multiple SPARK receptors instead of one. This can be seen in Fig. \ref{figmultiplespark1} that compares the data rate of one SPARK molecule with four with different chemical rates. The figure shows two plots having four curves; a red line indicates error probability for one SPARK, whereas the other curves represent the probability of error vs the data rate for $3, 5, 7$ number of molecules. Each set of curves is drawn for the same parameter, with the same number of observing states $n=2$ and time step $2$ms. The first figure shows the curves at low chemical rates, whereas the second set of curves is plotted for higher chemical rates. It can be seen that the error probability is significantly lower as the number of SPARK molecules increases. This is further illustrated in Fig. \ref{figmultiplespark2}, which shows a comparison of one receptor with the receptors $3,5,7, 14, 28$ in terms of error probability and total time.   We have used the same parameters as used in the case of a single receptor, that is, the time step of $2$ms and state $n=2$. The figure shows that there is a significant decrease in the probability of error as compared to a single receptor. 

\begin{figure}[h!]
\centering
\includegraphics[width=\columnwidth]{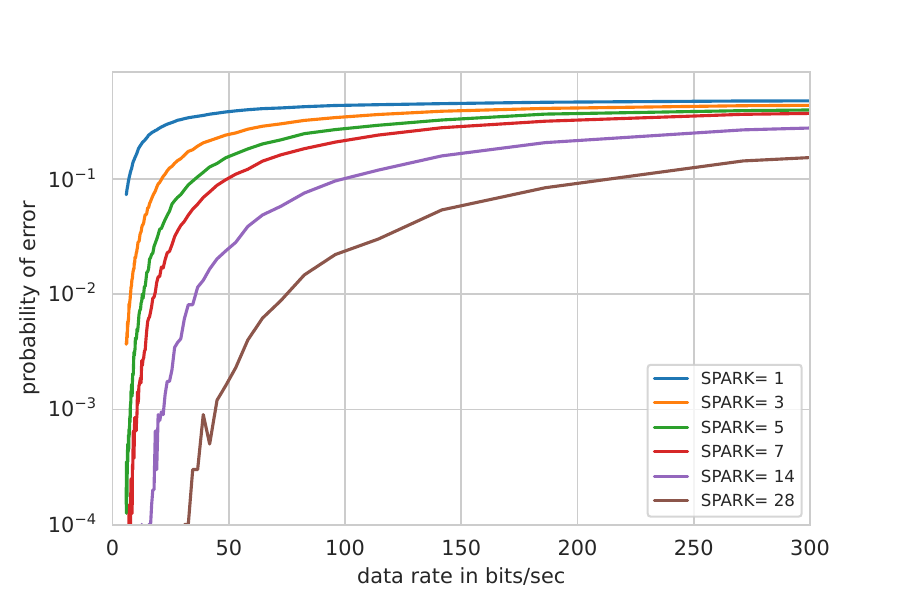}
\caption{Probability of error vs multiple SPARK receptors when the data rate is fixed. There is a dramatic impact on the probability of error based on the number of SPARKs. A single SPARK has a lower probability of error for certain data rates.}
\label{pberrvsdatamutiple}
\end{figure}

\begin{figure}[h!]
\centering
\includegraphics[width=\columnwidth]{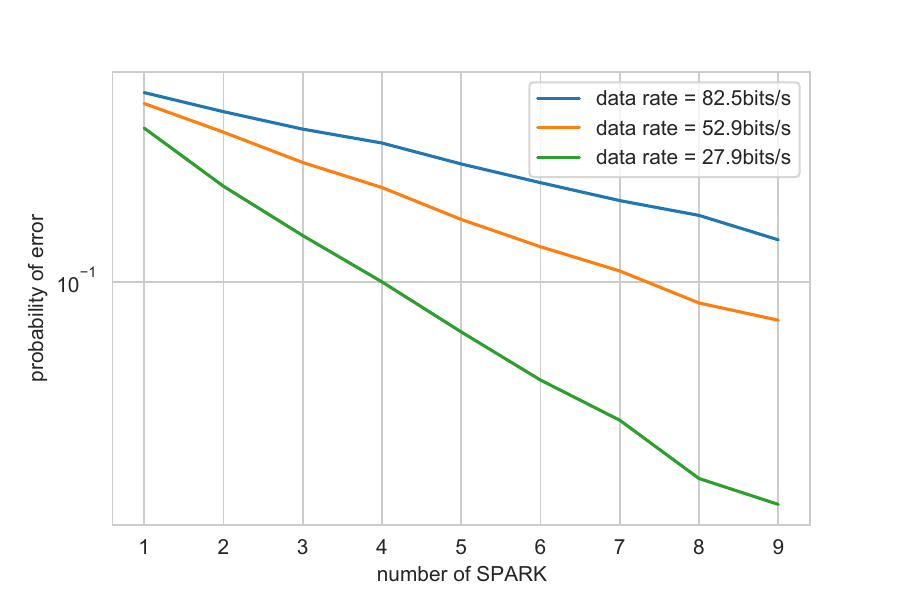}
\caption{Probability of error vs multiple SPARK receptors when the data rate is fixed. There is a dramatic impact on the probability of error based on the number of SPARKs. A single SPARK has a lower probability of error for certain data rates.}
\label{figfixedrate}
\end{figure}

The effect of using multiple receptors is further illustrated in Fig. \ref{figmultiplespark3}, Fig. \ref{pberrvsdatamutiple} and Fig. \ref{figfixedrate}. The curves of error probability against $\Delta t$ show $18\%$ less chances of error when the number of SPARK is $3$ than that of $1$ at  $\Delta t = 50$ms. The curves in Fig. \ref{pberrvsdatamutiple} show the same trend against data rates: the probability of error decreases as the number of receptors increases. This can be explained further in Fig. \ref{figfixedrate}, where we make a plot of error probability with respect to the number of SPARK receptors for fixed data rates. The curves show that the probability of error is higher for high data rates; however, as the number of SPARK increases, it becomes lower.\par

\textcolor{black}{Multiple SPARK receptors can be used to achieve higher data rates, for example, through a molecular multiple input multiple output (MIMO) technique \cite{channelrhodopsin2018}. Although the probability of error may be relatively high in some cases, error-correcting codes can be used to enhance reliability.}

\subsection{SPARK and Membrane Potential}
\rev{This section will explore the transduction property of synthetic photoisomerizable azobenzene-regulated K$^+$ (SPARK) to control a cell's membrane potential $V_m$. We used a simplified RC model (resistance-capacitance) to show the change of membrane potential according to the opening and closing of the SPARK channel. The control of $V_m$ is closely related to the charging and discharging behavior of the membrane capacitance, which is influenced by the state of the SPARK channels. When exposed to wavelength light around $380$ nm, the SPARK channels open, allowing potassium ions K$^+$ to flow out of the cell. This outflow of positive charge causes the membrane potential to decrease, effectively discharging the membrane capacitor as the potential becomes more negative. When the channels are exposed to visible light, they close, preventing the flow of K$^+$ ions. This cessation of ion flow allows the membrane potential to increase, similar to the charging of a capacitor, as the potential moves back towards a more positive value.}\par
\begin{figure}
     \centering
     \begin{subfigure}[b]{0.5\textwidth}
         \centering
         \includegraphics[width=\textwidth]{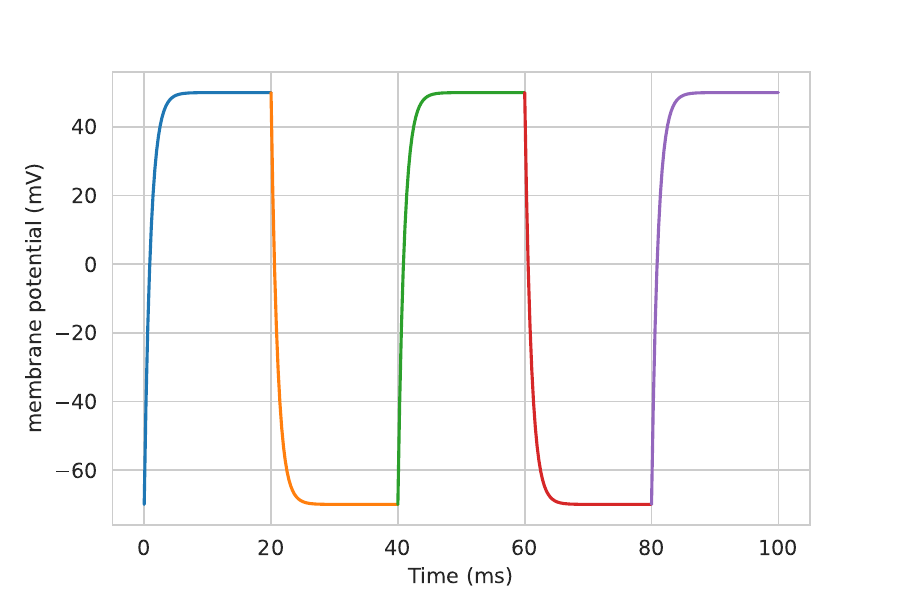}
         \caption{}
         \label{fig:membpotential1}
     \end{subfigure}
     \begin{subfigure}[b]{0.5\textwidth}
         \centering
         \includegraphics[width=\textwidth]{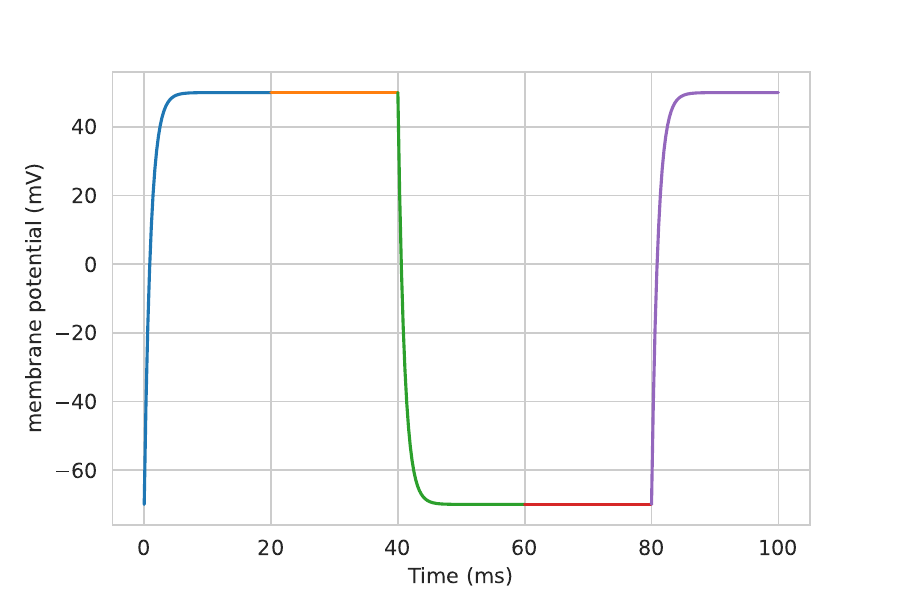}
         \caption{}
         \label{fig:membranepotential2}
     \end{subfigure}
        \caption{Charging of membrane potential with time in response of SPARK channel (a) when the bit sequence $[0,1,0,1,0]$ is transmitted. (b) when the bit sequence $[0,0,1,1,0]$. The membrane is simulated as RC circuit, with unit value of resistance and capacitance. The resting membrane potential $V_{rest}=-70mV$, while the membrane potential after charging is $V_{final}=50mV$. Interval of time $T = 20ms$ for each bit.}
        \label{fig:membpotentialcombined}
\end{figure}

\rev{To illustrate this, assume there is a sequence of bits $[0,1,0,1,0]$ which should be transmitted in a regular interval of time $T$, the bit $0$ will be transmitted by illuminating light of $380$nm, and the bit $1$ is transmitted by the wavelength of around $500$nm. The result of the simulation has been presented in Fig. \ref{fig:membpotentialcombined}. The resting membrane potential $V_{rest} = -70$ mV and the membrane potential after charging is $V_{final} = 50$mV. The typical value of the resting membrane potential in most cells, including neurons, ranges from $-60$ mV to $-90$ mV, with a common value being around $-70$ mV\cite{kandel2000principles}, \cite{khonsary2017guyton}. }

\section{Conclusion}
This study explores the feasibility of utilizing a single light-sensitive bio-receptor called SPARK for communication purposes. Growing biomolecules for use as receivers offers a sustainable and potentially scalable alternative to traditional manufacturing processes. Traditional electronic components are often made from rare and non-renewable materials. In contrast, biomolecules can be produced using renewable biological processes, such as fermentation and cell culture, which can be scaled up relatively easily. This approach reduces the environmental impact and resource dependency associated with conventional electronics manufacturing.The proposed system's concept is illustrated through a basic model comprising a transmitter (light source) and a receiver (SPARK), connected via free space, where information is conveyed through the illumination of the light source. The achieved data rates using this straightforward approach indicate that the proposed scheme holds promise as a potential candidate for future biomolecular communication systems.\par
\rev{In future research endeavors, enhancing the information rate could be achieved by incorporating multiple SPARK receptors. Additionally, transduction properties of SPARK can be explored in broader applications such as bioswitch and biosensor, where SPARK controls biological signals. Additionally, conducting information-theoretic analyses of the channel and communication system could offer insights to optimize information rates. This could involve assessing various parameters and configurations to maximize the efficiency and reliability of the communication process utilizing SPARK and similar bio-receptors.}
\appendix
\rev{Here, we provide the estimate of the light intensity required to initiate the transition from one state to another. The approximate value of energy required to transform from {\em trans} to {\em cis} is $200$ KJ/mol and $150-180$ KJ/mol \cite{beharry2011azobenzene},\cite{yager2006novel}.} 
In order to find out the energy per molecule, 
\begin{align}
 \label{eqn:energypermolecule}
    E_{molecule} &= \frac{200}{6.023 \times 10^{23}}\, ,\\
                 &= 3.32\times 10^{-19}.
\end{align}
where $6.023 \times 10^{23}$ is an Avogadro number representing the number of $molecules/mole$ 
Calculating the energy of a single photon using wavelength $\lambda = 300$ nm as
\begin{align}
    \label{eqn:energyperphoton}
    E_{photon} &= \frac{hc}{\lambda}\, ,\\
               &= \frac{6.626 \times 10^{-34}\cdot 3\times10^{8}}{300\times 10^{-9}}\, ,\\
               &= 6.626 \times 10^{-19} \:\text{J}.
\end{align}
From (\ref{eqn:energypermolecule}), we can see that the energy requirement for state transition is fulfilled from (\ref{eqn:energyperphoton}). The intensity $I$ of light is calculated as 
\begin{align}
    \label{eqn: intensity of light}
    I &= \frac{P}{A}\, ,\\
     P &= \frac{E}{t}\, ,\\
       &= \frac{3.32\times 10^{-19}}{20\times 10^{-3}}\, ,\\
       &= 1.66 \times 10^{-17} \:\text{W}\, ,\\
    I  &= \frac{1.66 \times 10^{-17}}{2\times 10^{-18}}\, \\
       &= 8.3 \:\text{W/m}^2\, .
\end{align}
In the above set of equations, $A$ is the area of the SPARK channel, which is roughly estimated on the basis of the size of a typical ion channel that is around $0.8$nm$^2$. The value of time $t$ is taken as $20$ms.
In a similar way we can calculate the intensity of light required for the transition from {\em cis} to {\em trans} using wavelength $\lambda = 500$nm. While the theoretical minimum intensity requirement is $7-8$ W/m$^2$, we used 100 W/m² in simulations to ensure robust transition and reliable system performance.


\bibliographystyle{ieeetr}
\bibliography{globecom21.bib}
\begin{IEEEbiography}[{\includegraphics[width=1in,height=1.25in, clip]{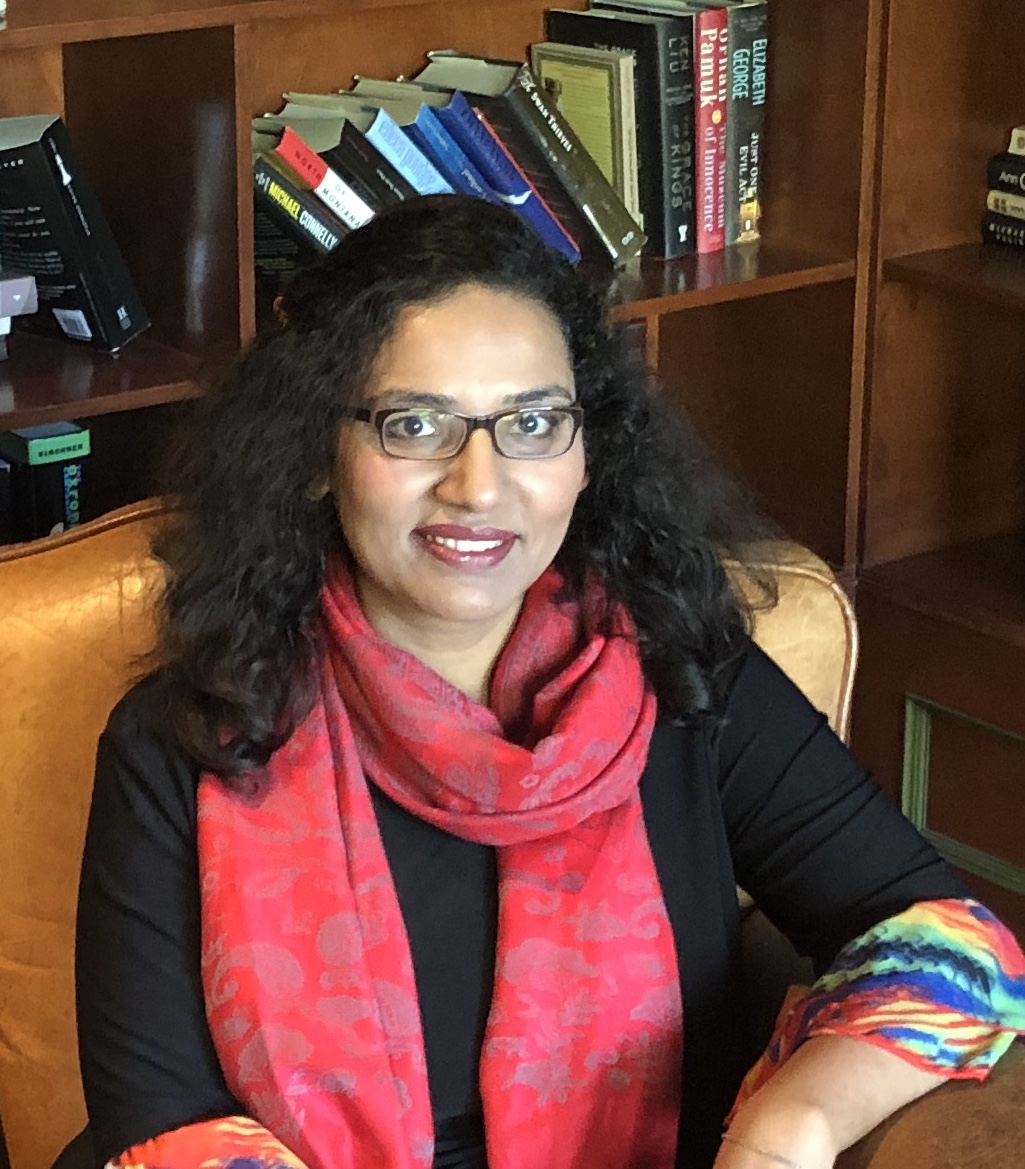}}]{Taha Sajjad} (Member, IEEE) recently earned her Ph.D. in Electrical and Computer Engineering from the Lassonde School of Engineering at York University. Currently, she is a researcher in Prof. Andrew Eckford’s Lab, where she explores cutting-edge applications of molecular communication. Her work bridges multiple disciplines, including communication systems, embedded electronics, and biomedical instrumentation. Beyond research, Taha actively engages in science outreach initiatives such as Soapbox Science 2024, and she has translated a comprehensive text on mind sciences that highlights the transformative power of human cognition. She also serves as a peer reviewer for the IEEE Vehicular Technology Conference (VTC) 2025 and has made contributions to graduate student leadership and volunteer efforts within the academic community.

\end{IEEEbiography}
\begin{IEEEbiography}[{\includegraphics[width=1in,height=1.25in,clip]{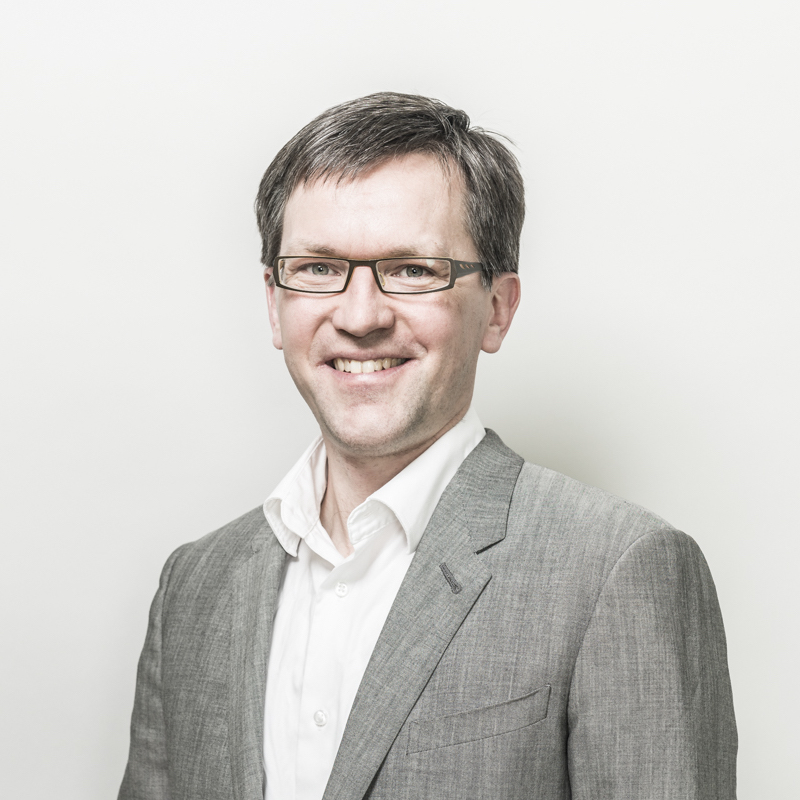}}]{Andrew W. Eckford} (Senior Member, IEEE) received the B.Eng. degree in electrical engineering from the Royal Military College of Canada in 1996, and the M.A.Sc. and Ph.D. degrees in electrical engineering from the University of Toronto in 1999 and 2004, respectively. He was a Postdoctoral Fellowship with the University of Notre Dame and the University of Toronto, prior to taking up a faculty position with York, in 2006. He is an Associate Professor with the Department of Electrical Engineering and Computer Science, York University, Toronto, ON, Canada. He has held courtesy appointments with the University of Toronto and Case Western Reserve University. In 2018, he was named a Senior Fellow of Massey College, Toronto. He is also a coauthor of the textbook Molecular Communication (Cambridge University Press). His research interests include the application of information theory to biology and the design of communication systems using molecular and biological techniques. His research has been covered in media, including The Economist, The Wall Street Journal, and IEEE Spectrum. His research received the 2015 IET Communications Innovation Award, and was a Finalist for the 2014 Bell Labs Prize.
\end{IEEEbiography}


\end{document}